\documentclass[pre,aps,twocolumn,superscriptaddress,floatfix]{revtex4-1}

\usepackage{amssymb}
\usepackage{epsfig}
\usepackage{amsmath}
\usepackage{subfigure}
\usepackage{graphicx}
\usepackage{textcomp}
\usepackage{url}
\usepackage{float}
\usepackage{color}
\usepackage{cases}
\usepackage[normalem]{ulem}
\usepackage[titletoc,title]{appendix}

\usepackage[colorlinks=true, urlcolor=blue, anchorcolor=blue, citecolor=blue,filecolor=blue,linkcolor=blue,menucolor=blue
]{hyperref}

\usepackage{color}

\usepackage{fixmath}

\begin{document}
\title{Competition in a system of Brownian particles: Encouraging achievers}

\author{P. L. Krapivsky}
\email{pkrapivsky@gmail.com}
\affiliation{Department of Physics, Boston University, Boston, Massachusetts 02215, USA}
\affiliation{Santa Fe Institute, Santa Fe, New Mexico 87501, USA}
\author{Ohad Vilk}
\email{ohad.vilk@mail.huji.ac.il}
\affiliation{Racah Institute of Physics, Hebrew University of Jerusalem, Jerusalem 91904, Israel}
\affiliation{Movement Ecology Lab, Department of Ecology, Evolution and Behavior, Alexander Silberman Institute of Life Sciences, Hebrew University of Jerusalem, Jerusalem 91904, Israel}
\affiliation{Minerva Center for Movement Ecology, Hebrew University of Jerusalem, Jerusalem 91904, Israel}
\author{Baruch Meerson}
\email{meerson@mail.huji.ac.il}
\affiliation{Racah Institute of Physics, Hebrew University of Jerusalem, Jerusalem 91904, Israel}

\begin{abstract}
We introduce and study analytically and numerically a simple model of inter-agent competition, where underachievement is strongly discouraged. We consider $N\gg 1$ particles performing independent Brownian motions on the line. Two particles are selected at random and at random times, and the particle closest to the origin is reset to it. We show that, in the limit of $N\to \infty$, the dynamics of the coarse-grained particle density field can be described by a nonlocal hydrodynamic theory which was encountered in a study of the spatial extent of epidemics in a critical regime. The hydrodynamic theory predicts relaxation of the system toward a stationary density profile of the ``swarm" of particles, which exhibits a power-law decay at large distances. An interesting feature of this relaxation is a non-stationary ``halo" around the stationary solution, which continues to expand in a self-similar manner. The expansion is ultimately arrested by finite-$N$ effects at a distance of order $\sqrt{N}$ from the origin, which gives an estimate of the average radius of the swarm. The hydrodynamic theory does not capture the behavior of the particle farthest from the origin  -- the current leader. We suggest a simple scenario for typical fluctuations of the leader's distance from the origin and show that the mean distance continues to grow indefinitely as $\sqrt{t}$.  Finally, we extend the inter-agent competition from $n=2$ to an arbitrary number $n$ of competing Brownian particles ($n\ll N$). Our analytical predictions are supported by Monte-Carlo simulations.
\end{abstract}

\maketitle

\section{Introduction}

Recent years have witnessed a significant interest in stochastic reset models. Motivated by optimization of random search, a basic reset model was introduced in Ref.~\cite{EM2011}: A single Brownian particle on the line that is stochastically reset to a specified point. In this model, a target is found in a finite time, in contrast to Brownian motion without reset \cite{rednerbook}.  Apart from the random search optimization, the basic reset model is interesting because it exhibits a simple nonequilibrium steady state (NESS) \cite{EM2011}: a convenient platform for probing different aspects of statistical mechanics out of equilibrium. The reset model \cite{EM2011} has been extended to many other stochastic processes and settings, see Ref.~\cite{EMS2020} for a recent review.

When there is a population of $N\gg 1$ particles subject to random resets, the problem acquires qualitatively new features (``more is different") and becomes quite rich, especially in the presence of inter-particle interactions \cite{VAM2022}. Furthermore, $N$-particle reset models have close relatives among the family of Brunet-Derrida $N$-particle models: branching Brownian motions with selection \cite{BD1,BD2}. In these models when a branching event occurs, the particle with the lowest fitness is eliminated. The Brunet-Derrida $N$-particle models differ by the choice of the fitness function mimicking different circumstances of biological selection \cite{BD3,BD4,BD5,BD6,BD7,BD8,bees1,bees2,bees3,bees4,bees5}. The ``Brownian bees" model \cite{bees1,bees2,bees3,bees4,bees5} is especially similar to $N$-particle models with reset. In the Brownian bees model, the particle farthest from the origin \cite{bees1,bees2,bees4,bees5} (or from the instantaneous center of mass of the system \cite{bees3}), is eliminated immediately when a branching  event occurs. The similarity becomes evident upon an observation that the Brownian bees model can be reformulated as a reset model. Indeed, a simultaneous process of branching and elimination of the farthest particle is equivalent to resetting the farthest particle to the location of any of the remaining $N-1$ particles.

Reset of the farthest particle encapsulates (i) global competition among all the particles and (ii) discouragement of achievers.  In this paper, we modify the competition rules significantly. First, the competition is now among \emph{two} randomly chosen particles (later on, we will extend it to an arbitrary number $n\ll N$ of particles). Second, the competition  discourages underachievers: it is the competitor \emph{closest} to the origin which is reset to the origin and should start from scratch.

Our motivation to study this model is partly due to the widely known fact that wealth is not distributed equally across society. As early as in 1897, Pareto showed that the distribution of incomes follows a power law \cite{Pareto1897}, a finding that has since been validated in many studies, see e.g., \cite{Solomon1997New,forbes2006}. 
Our toy model is based on the idea of nonlocal competition and reset, and it leads to the emergence of a broad wealth distribution. We make no pretense of accounting for the many real factors leading to inequality of wealth. However, as we show below, our model does capture some prominent features of wealth distribution in a society in the limit of many interacting agents.

We will show that, in the limit of $N\to \infty$, the behavior of all particles except for the current leader (see below) can be described by a nonlocal hydrodynamic theory. This theory can be brought to the form of a reaction-diffusion equation which was encountered previously in multiple contexts. The hydrodynamic theory predicts how the coarse-grained particle density field approaches a steady state which exhibits a power-law tail at large distances. This is in contrast with the previously studied Brownian reset models \cite{VAM2022,bees1,bees2,bees3}, where achievement is discouraged to such an extent that the steady-state hydrodynamic density profile has a compact support. In addition, the relaxation of the coarse-grained density to a steady state exhibits a non-stationary ``halo" around the stationary solution. This halo expands in a self-similar manner until the expansion is arrested, due to finite-$N$ effects, at a distance of order $\sqrt{N}$ from the origin.

The dynamics of the leader, defined as the particle which is, at a given time, farthest from the origin, is entirely different. The leader (whose identity changes in time) never loses in the competition, so it continues its Brownian explorations forever. In particular, its average distance from the origin increases indefinitely with time as $\sqrt{t}$. We suggest a simple scenario for typical fluctuations of the leader's position at long times. In this scenario, (i) the leader performs Brownian motion, (ii) the rest of the particles form the steady-state swarm, and (iii) the average radius of the swarm  serves as a reflecting wall for the leader.  We verify this and other main analytical predictions in Monte-Carlo simulations.

The remainder of the paper is organized as follows. In Sec.~\ref{sec:2} we present the hydrodynamic model for the coarse-grained particle density and obtain the steady state density profile. In Sec.~\ref{relaxation} we study the relaxation of the density profile toward the steady state at long times. Section~\ref{finiteN} deals with an important finite-$N$  effect: the arrest of the swarm's expansion. In Sec.~\ref{leader} we present our simple scenario for the dynamics of the current leader, and in Sec. \ref{sec:n} we extend the competition to an arbitrary number of $n$ particles (we require $n\ll N$).  We summarize and discuss our results, in particular with respect to the distribution of wealth and inequality, in Sec.~\ref{discussion}.

\section{Hydrodynamic theory and steady state}
\label{sec:2}

We restate the rules of our model for clarity: $N$ Brownian particles perform Brownian motion on the line. At a constant reset rate, two particles are randomly chosen, and the one closest to the origin $x=0$ is reset to it. We rescale time and distance so that the diffusion constant and the reset rate per particle are both equal to 1.

When $N\gg 1$, the dynamics of all the particles except the leader (the particle currently farthest from the origin) can be described by the hydrodynamic theory which ignores fluctuations. The theory has the form of a nonlocal equation in partial derivatives for the coarse-grained particle density $\rho(x,t)$, rescaled by $N$.  In addition to the conventional diffusion term, the governing equation accounts for the particle's reset due to the pairwise competition. We obtain
\begin{eqnarray}
  \rho_t \!&=& \!\rho_{xx} - 4 \rho \left[\int \displaylimits_{-\infty}^{-|x|}\rho(y,t) dy  + \int\displaylimits_{|x|}^{\infty}\rho(y,t) dy \right]\nonumber \\
  \! &+& \!4 \delta(x) \!\!\! \int\displaylimits_{-\infty}^{\infty}\!dy\,\rho(y,t)\!\!\left[\int\displaylimits_{-\infty}^{-|y|}\!dz\,\rho(z,t)
   +\int\displaylimits_{|y|}^{\infty}\!dz\,\rho(z,t)\right].
  \label{eqrhoallaxis}
\end{eqnarray}
The second term on the right describes the particle loss at position $x$. This term represents, in the continuous limit, the total number of pairs of particles at positions $x$ and $y$ such that $|y|<|x|$. The factor $4$ in the loss terms assures that the  total rate of particle reset, rescaled by $N$, is equal to the unity. The last term on the right describes the reappearance of the reset  particles at the origin. This term is determined by the condition that the total number of particles in the system is conserved at all times.

Let us assume for simplicity that the initial particle density is symmetric, $\rho(x,t=0)=\rho(-x,t=0)$. Then $\rho(x,t)$ remains symmetric throughout the evolution, and Eq.~\eqref{eqrhoallaxis} can be rewritten as
\begin{equation}
  \rho_t = \rho_{xx} - 4 \rho \int_x^{\infty}\rho(y,t) dy\,, \quad x>0\,.
  \label{eqrho}
\end{equation}
A greater simplification, however, is obtained upon transforming Eq.~\eqref{eqrho} into an equation for the fraction of particles on the $(x,\infty)$ ray,
\begin{equation}
\label{r:def}
r(x,t) = \int_x^\infty dy\,\rho(y,t)\,.
\end{equation}
Indeed, plugging this definition into Eq.~(\ref{eqrho}) and integrating over $x$ we obtain
\begin{equation}
\label{r-2}
r_t = r_{xx}-2 r^2\,.
\end{equation}
It is evident from Eq.~(\ref{r:def}) that
\begin{equation}
\label{r:BC}
r(0,t)=\frac{1}{2} \quad \text{and}\quad r(\infty,t)=0,
\end{equation}
and we can consider Eq.~(\ref{r-2}) on the $x>0$ half-line with the boundary conditions (\ref{r:BC}) and a specified initial condition for $r(x,t=0)$.

Equation \eqref{r-2} is a reaction-diffusion equation \cite{Smoller,Evans}. If one interprets $r(x,t)$ as the particle density, Eq.~\eqref{r-2} provides a mean-field description to aggregation and annihilation processes with diffusing reactants \cite{OTB89,book}. This equation and its stationary solution, see below, have also appeared in other contexts, see \textit{e.g.} Refs.~\cite{allele,extent,Mwall}.

\begin{figure}[t]
\includegraphics[width=0.35\textwidth,clip=]{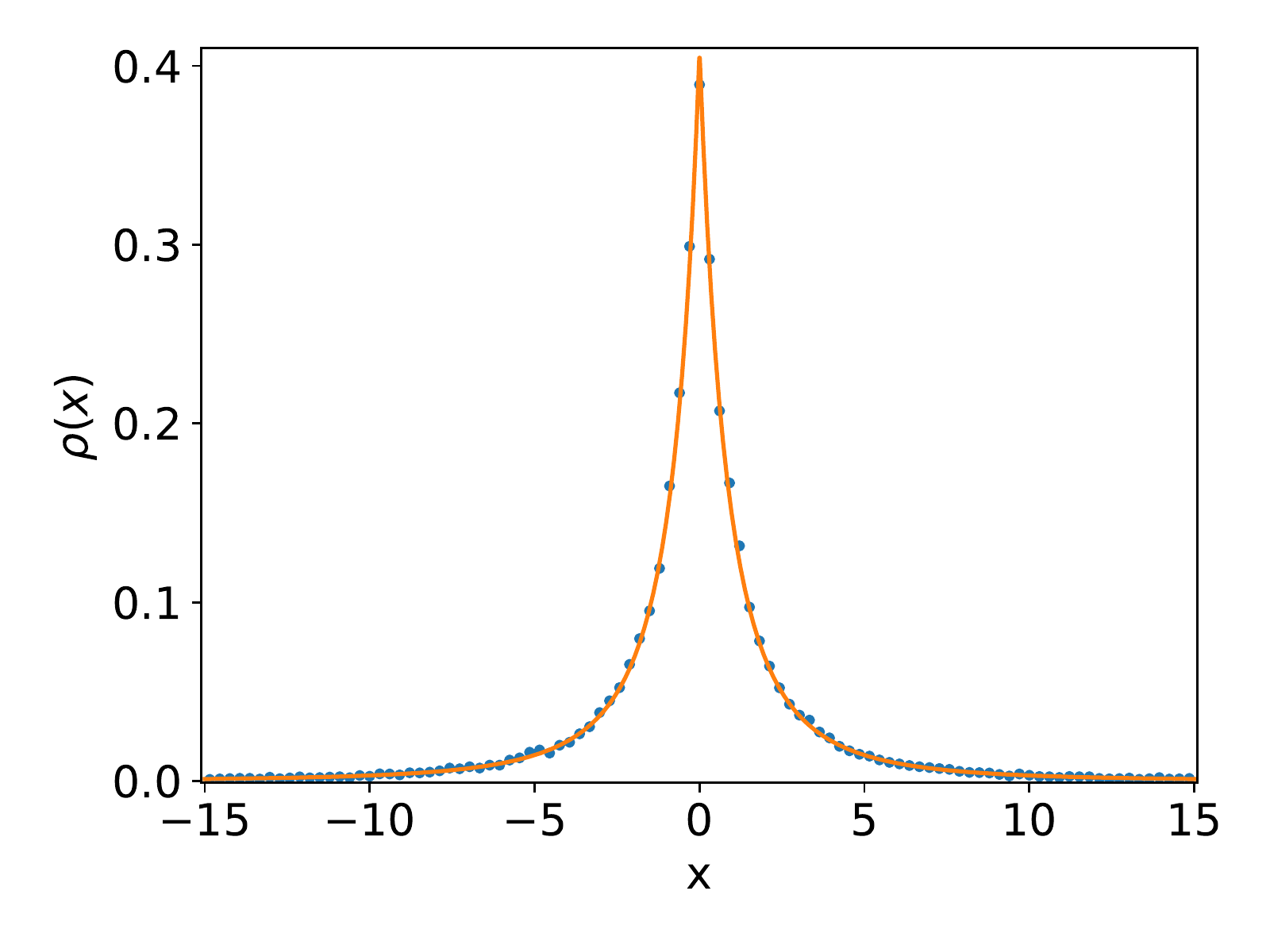}
\caption{The steady state density profile for pair competition with $N=10^2$ particles, as observed in simulations (points) and
predicted by Eq.~\eqref{rho-2:sol} (solid line).}
\label{fig:rho2}
\end{figure}

The steady state solution of Eq.~\eqref{r-2} solves the equation
\begin{equation}\label{steady}
r''-2r^2=0\,,
\end{equation}
where the primes denote the $x$ derivatives. The ``energy integral" of Eq.~(\ref{steady}) can be written as
\begin{equation}\label{steady1}
(r')^2-\frac{4}{3}r^3=0\,,
\end{equation}
where the constant in the r.h.s. is zero by virtue of the second boundary condition in Eq.~(\ref{r:BC}).
The proper solution of Eq.~(\ref{steady1}) is
\begin{equation}
\label{r2:sol}
r_0(x) = \frac{1}{2}\left(1+\frac{x}{\sqrt{6}}\right)^{-2}\,,\quad x>0,
\end{equation}
where we used the first boundary condition in Eq.~(\ref{r:BC}).
Differentiating Eq.~(\ref{r2:sol}) with respect to $x$, we arrive at the steady-state density
\begin{equation}
\label{rho-2:sol}
\rho_0(x) = \frac{1}{\sqrt{6}}\left(1+\frac{|x|}{\sqrt{6}}\right)^{-3}\,, \quad |x|<\infty\,,
\end{equation}
where we have taken into account the $x\leftrightarrow -x$ symmetry. The main feature of the  steady-state (\ref{rho-2:sol}) is the power-law tail $\sim |x|^{-3}$ at $|x|\gg 1$. This is in contrast with the previously studied Brownian reset models \cite{VAM2022,bees1,bees2,bees3}, where achievement is discouraged to such an extent that the steady-state hydrodynamic density profile has a compact support. The corner singularity (a jump in the first derivative) at $x=0$ is a direct consequence of the presence of the delta-function source at $x=0$, see Eq.~\eqref{eqrhoallaxis}; it is a common feature of models where the particles are reset to a single point.   Equation~\eqref{rho-2:sol} is in good agreement with our Monte-Carlo simulations [see Fig.~\ref{fig:rho2}].

\section{Relaxation to steady state}
\label{relaxation}

The relaxation of the density field $\rho(x,t)$ to the stationary solution \eqref{rho-2:sol} is described by the time-dependent equation~\eqref{r-2}, and it is nontrivial \cite{extent}. At long times,  $t\gg 1$, the relaxation process obeys the remarkable asymptotic scaling solution  \cite{extent}
\begin{equation}
\label{rxt}
r(x,t) = r_0(x)\mathcal{R}(\xi)\,,
\end{equation}
where $\xi=x/\sqrt{4t}$ is the scaling variable. The ansatz (\ref{rxt}) describes the
establishment of the stationary solution $r_0(x)$ in an expanding region of space surrounded by a non-stationary halo which is expanding in a self-similar manner.

The scaling function $\mathcal{R}(\xi)$ has not been determined previously. To compute it, we insert the ansatz  \eqref{rxt} into Eq.~\eqref{r-2}. In the limit of $t\to \infty$ this leads to an ordinary differential equation
\begin{equation}
\label{R:eq}
\mathcal{R}''(\xi)+\left(2\xi-\frac{4}{\xi}\right)\mathcal{R}'(\xi)
+\frac{6}{\xi^2}\,\mathcal{R}(1-\mathcal{R})=0\,.
\end{equation}
The boundary conditions [cf. Eqs.~\eqref{r:BC} and \eqref{rxt}] are
\begin{equation}
\label{R:BC}
\mathcal{R}(0) = 1, \qquad \mathcal{R}(\infty)=0\,.
\end{equation}
The nonlinear problem (\ref{R:eq}) and (\ref{R:BC}) is parameter-free, and we solved it numerically by a shooting method.  $\xi=0$ is a singular point of Eq.~(\ref{R:eq}), and the existence of a regular solution demands that the first and second derivatives of $R$ vanish at $\xi=0$. The asymptotic of $R(\xi)$ at $\xi \to 0$ can be found perturbatively. We set $R(\xi) = 1-u(\xi)$, where $\xi\ll 1$ and $u(\xi) \ll 1$. To leading order the equation for $u(\xi)$ is the following \begin{equation}
\label{R:small}
u''-\frac{4}{\xi} u'-\frac{6}{\xi^2} u=0.
\end{equation}
The solution is a linear combination of $\xi^6$ and $\xi^{-1}$. The $\xi^{-1}$ term must be ruled out, and we obtain
\begin{equation}\label{R:small1}
\mathcal{R}(\xi \ll 1) = 1- A \xi^6+ \dots \,\quad \xi\ll 1 ,
\end{equation}
where $A=O(1)$ is an a priori unknown constant. Interestingly, not only the first and second derivatives, but also the third, fourth and fifth derivatives of $R(\xi)$ vanish at $\xi=0$.

At $\xi\gg 1$ Eq.~(\ref{R:eq}) simplifies to
\begin{equation}
\label{R:largeeq}
\mathcal{R}''(\xi)+2\xi  \mathcal{R}'(\xi)=0.
\end{equation}
The solution vanishing at infinity behaves as
\begin{equation}\label{R:largesol}
\mathcal{R}(\xi) \simeq \frac{B\, e^{-\xi^2}}{\xi}\,.
\end{equation}
A unique value of the coefficient $B=O(1)$, for which the asymptotic (\ref{R:largesol}) matches with the ``body" of $R(\xi)$, can only be found numerically.

\begin{figure}[ht]
\includegraphics[width=0.35\textwidth,clip=]{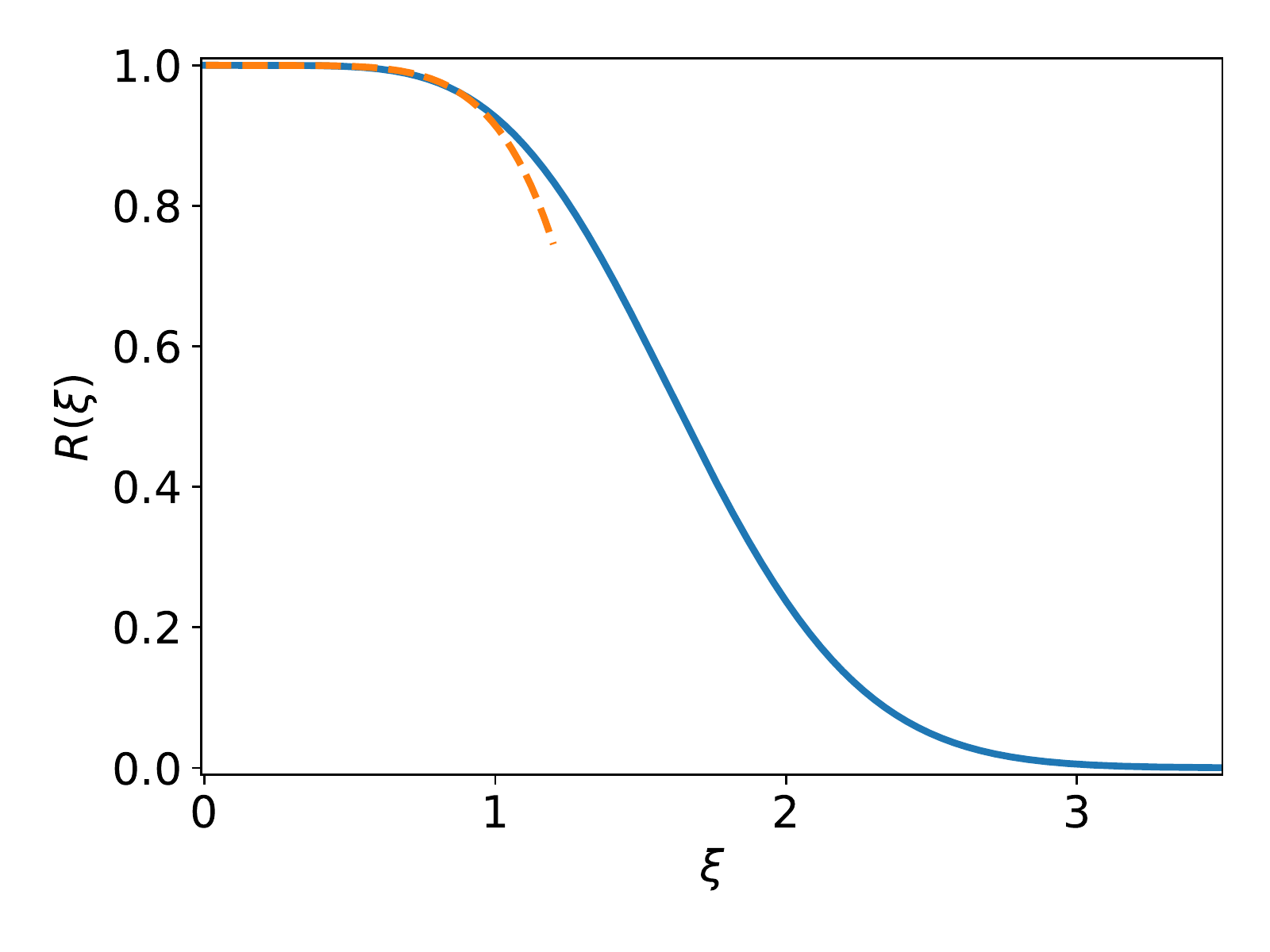}
\caption{The shape function $R(\xi)$, see Eq.~\eqref{rxt}, determined by solving Eqs.~(\ref{R:eq}) and (\ref{R:BC}) numerically. The dashed line shows the small-$\xi$ asymptotic (\ref{R:small1}).
}
\label{Rxi}
\end{figure}

Using the small-$\xi$ asymptotic (\ref{R:small1}), we solved the Cauchy problem
for Eq.~(\ref{R:eq}) on a finite interval $(\epsilon,L)$, where $0<\epsilon \ll 1$. The numerical solution
approaches zero at large $\xi$ for a single value of the constant $A$, and this constant was used as the shooting parameter. Figure \ref{Rxi} shows the resulting shape function $\mathcal{R}(\xi)$ alongside with the asymptotic \eqref{R:small1}.

\begin{figure}[ht]
\includegraphics[width=0.35\textwidth,clip=]{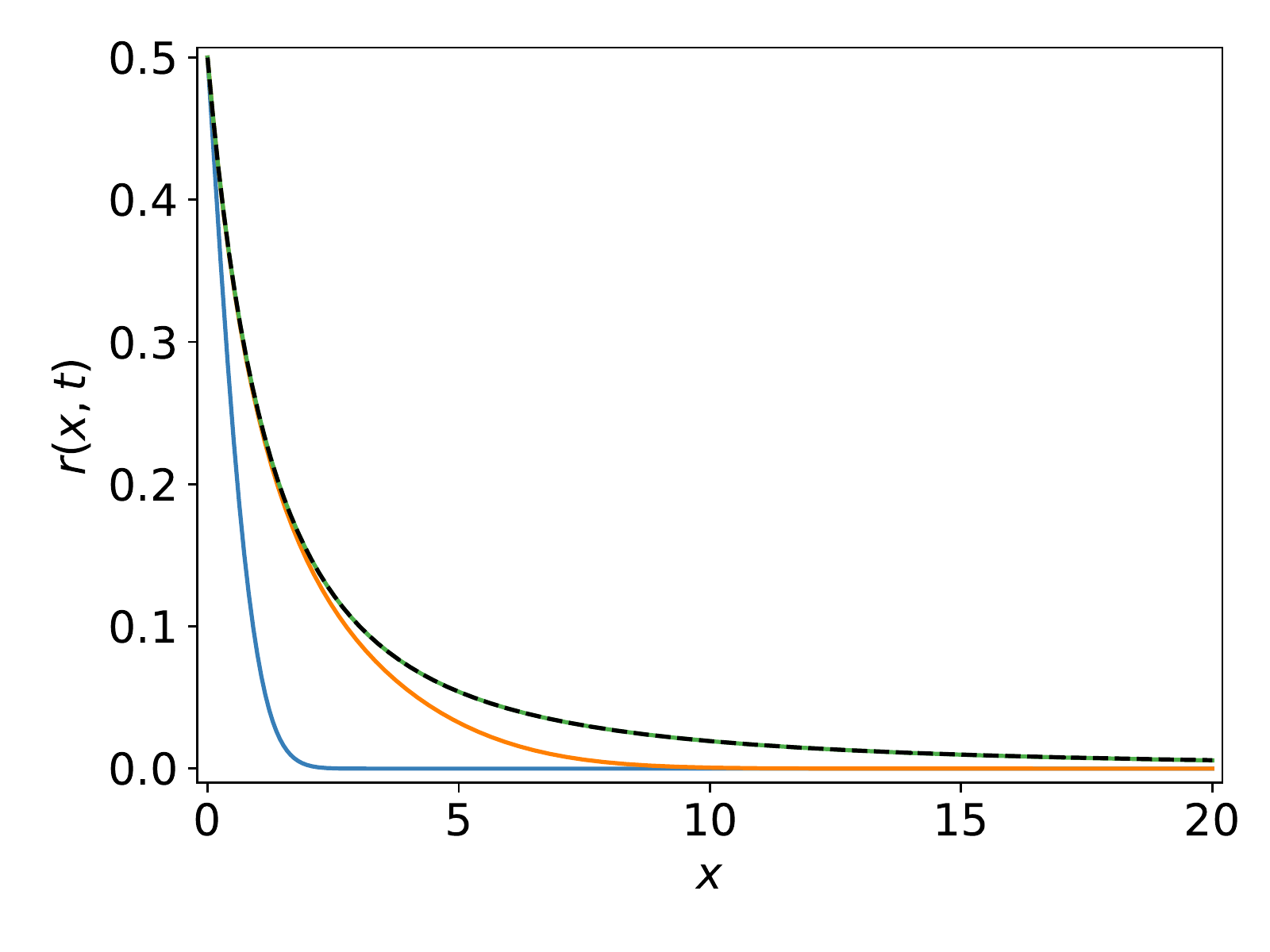}
\caption{The time-dependent numerical solution for $r(x,t)$ approaches the steady state solution $r_0(x)$. Solid lines: $r(x,t)$ at times  $0$, $5$, $200$ and $400$ (from left to right). The last two lines coincide with each other and with the analytical prediction (\ref{r2:sol}) for the steady state (dashed line).}
\label{pde1}
\end{figure}

\begin{figure}[ht]
\includegraphics[width=0.35\textwidth,clip=]{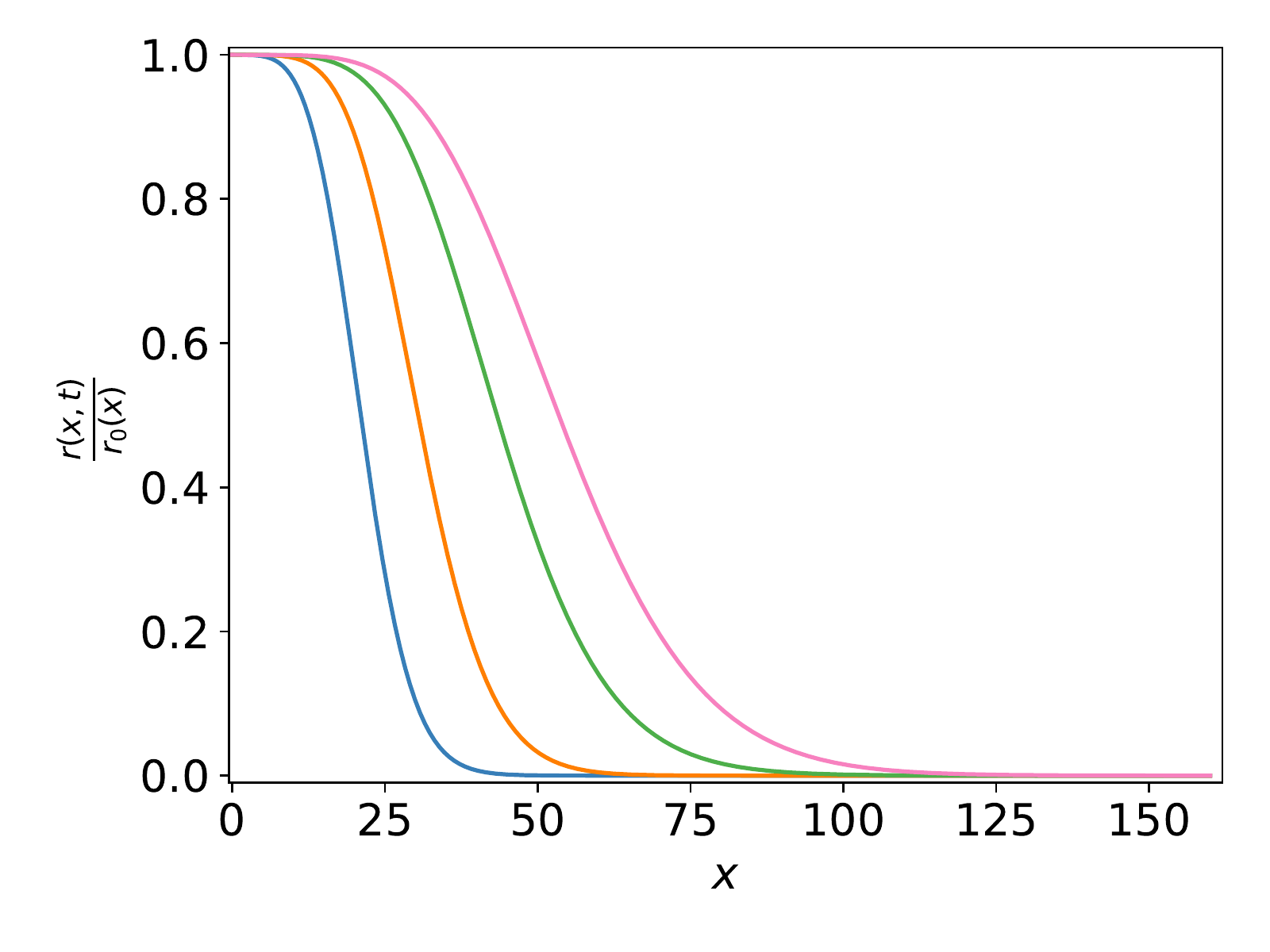}
\caption{The time-dependent numerical solution for $r(x,t)$ exhibits an expanding halo as predicted by Eq.~(\ref{rxt}). Shown is the ratio $r(x,t)/r_0(x)$ at times $50$, $100$, $200$ and $400$ (from left to right).}
\label{pde2}
\end{figure}

\begin{figure}[ht]
\includegraphics[width=0.5\textwidth,clip=]{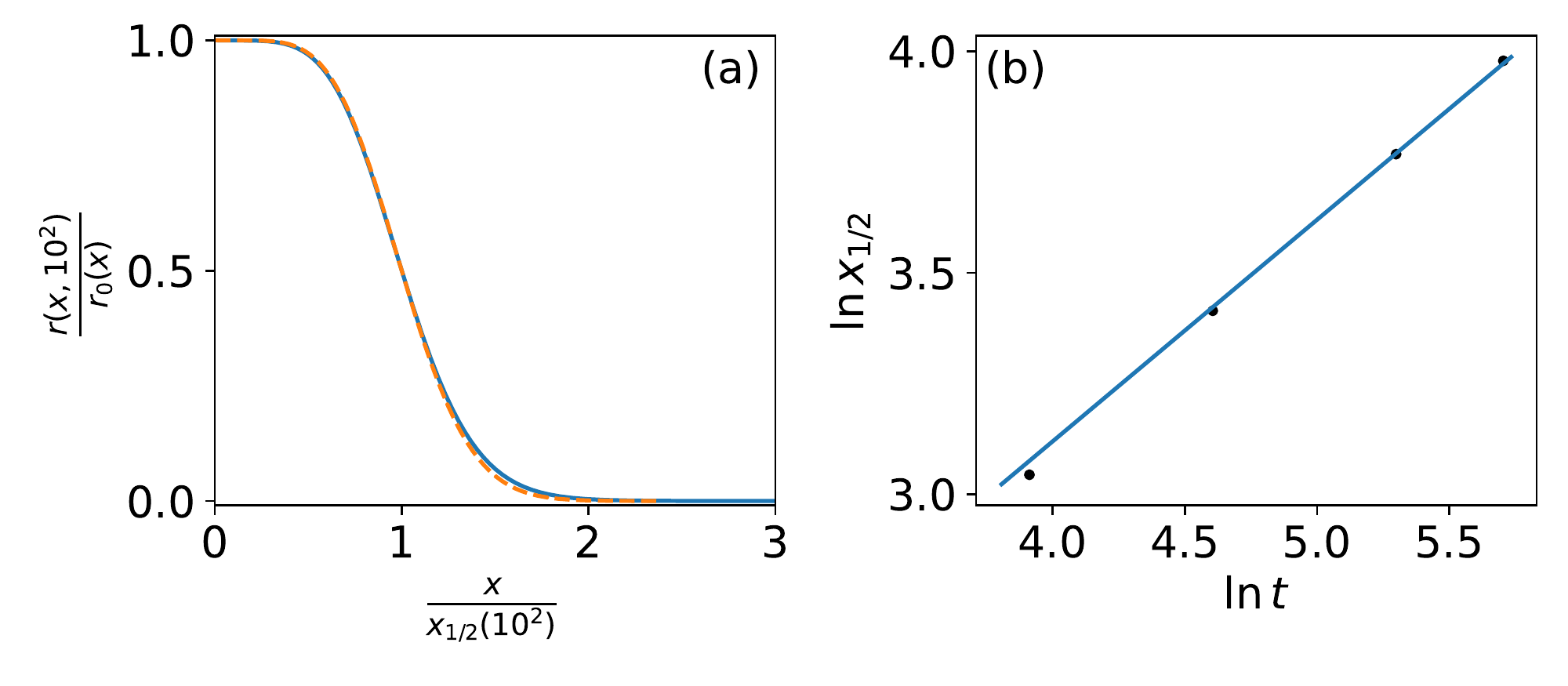}
\caption{(a) The ratio $r(x,t=100)/r_0(x)$ (solid line) is compared with the shape function $R(\xi)$ from Eq.~(\ref{R:eq}) (dashed line). The horizontal axis is rescaled to the distance $x_{1/2}(t)$ where the depicted functions are both equal to $1/2$. (b) $\ln x_{1/2}(t)$ versus $\ln t$ at $t=50$, $100$, $200$ and $300$ (symbols). A fit (straight line) gives the dynamic exponent $0.51$ in agreement with Eq.~(\ref{rxt}).}
\label{pde3}
\end{figure}

To verify the long-time solution (\ref{rxt}), we solved numerically the time-dependent Eq.~\eqref{r-2} with boundary conditions \eqref{r:BC} and a localized initial density. Figure \ref{pde1} shows how the numerical solution relaxes to the steady state (\ref{r2:sol}). Figure \ref{pde2} provides a closer look at the solution at larger distances by showing how the halo of $r(x,t)$ expands in time. Finally,
Fig.~\ref{pde3} compares, at $t=100$, the ratio $r(x,t)/r_0(x)$  with the shape function $\mathcal{R}(\xi)$, found from Eq.~(\ref{R:eq}), and also verifies the dynamical scaling $x \sim \sqrt{t} $.

To appreciate the nontrivial character of the system's long-time relaxation, let us consider the density moments for positive $x$:
\begin{equation}
\label{Ma:def}
M_a(t) = \langle x^a\rangle = \int_0^\infty dx\, x^a \rho(x,t)\,,
\end{equation}
for arbitrary  $a \geq 0$. For $a<2$ the presence of the expanding halo is inconsequential in the leading order. Indeed, plugging $\rho(x,t)=r_0(x)$ into Eq.~\eqref{Ma:def}, we obtain
\begin{equation}\label{Mastatic}
M_a = 6^{a/2}\frac{\pi a(1-a)}{2\sin(\pi a)}\,,
\end{equation}
This expression is finite for $a<2$, and it is independent of time. As to be expected, $M_0=1/2$.

For $a\geq 2$ the $a$-moments of $\rho_0(x)$ diverge. A finite result for $M_a(t)$, increasing indefinitely with time, appears because of the halo, which causes the integral to converge.  In the marginal case $a=2$ the exact form of the shape function $\mathcal{R}(\xi)$ is unimportant to logarithmic accuracy \cite{extent}. Indeed, using the relation $\rho(x,t)=-r_x(x,t)$ and \eqref{rxt}, one obtains
\begin{eqnarray}
M_2(t) &=&  \int_0^\infty dx\, x^2 \rho(x,t)= 2\int_0^\infty dx\, x \,r(x,t) \nonumber \\
& = & 2\int_0^\infty dx\, x\, r_0(x)\mathcal{R}\left(\frac{x}{\sqrt{4t}}\right) \nonumber \\
& \simeq &6 \int_1^{\sqrt{t}} \frac{dx}{x} = 3\ln t,
\label{logscaling}
\end{eqnarray}
a logarithmic scaling with time. This result was obtained in Ref.~\cite{extent}.

For $a>2$, $M_a$ grows with time as a power law, and the numerical coefficient of the power law explicitly depends on the scaling function $\mathcal{R}(\xi)$. Overall, we obtain
\begin{equation}
M_a(t) \simeq
\begin{cases}
6^{a/2}\frac{\pi a(1-a)}{2\sin(\pi a)}\,, & a<2\,,\\
3\ln t\,,                                                & a=2\,,\\
m_a (4t)^{\frac{a}{2}-1}\,,                             & a>2\,,
\end{cases}
\label{threeeq}
\end{equation}
where
\begin{equation}
\label{R-a}
m_a = 3a\int_0^\infty d\xi\,\xi^{a-3} \mathcal{R}(\xi)\,.
\end{equation}
A numerical evaluation gives $m_3\simeq 15$, $m_4\simeq 18$, and $m_5\simeq 29$.

\begin{figure}[t]
\includegraphics[width=0.35\textwidth,clip=]{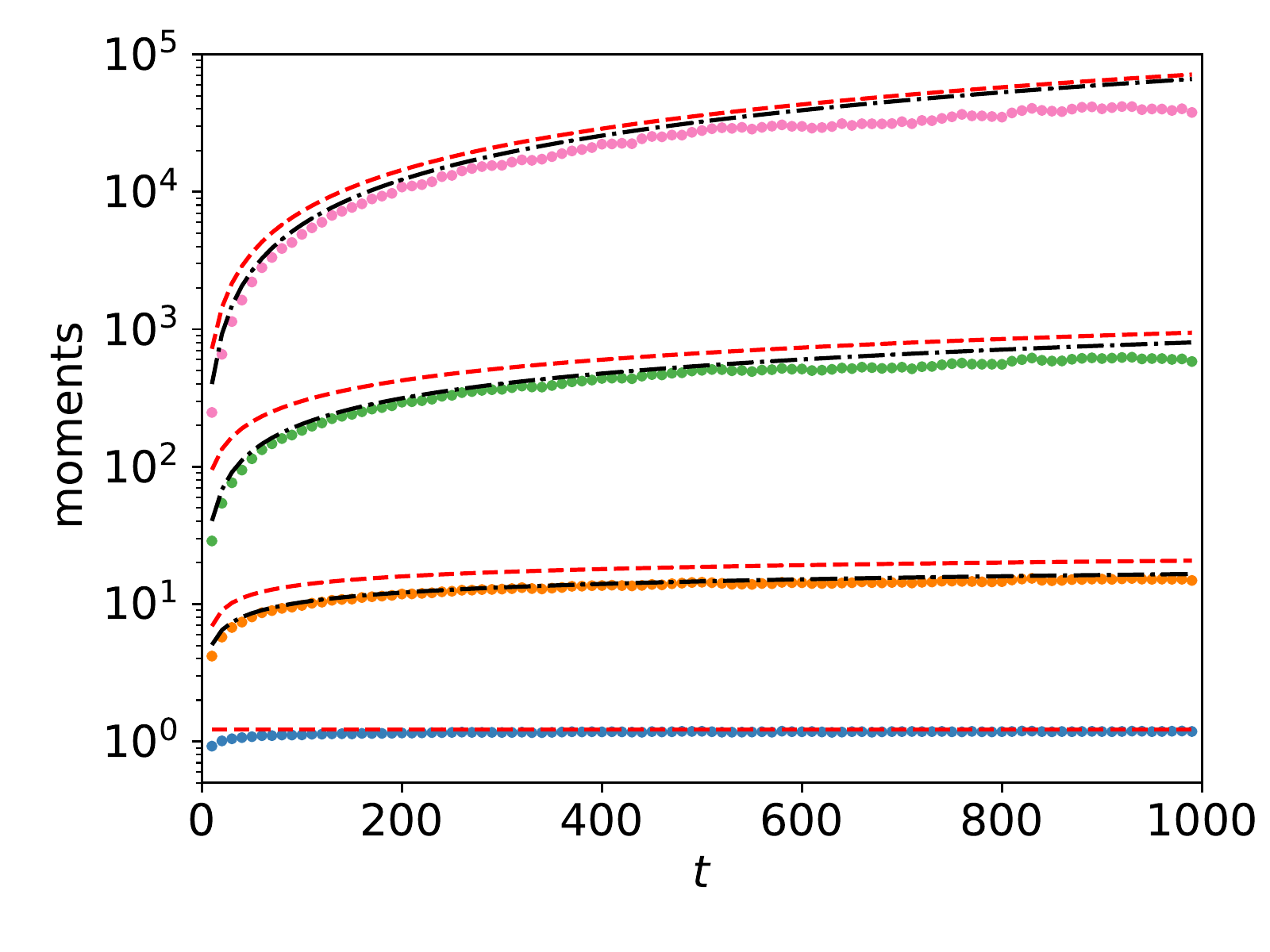}
\caption{The density moments \eqref{Ma:def} versus time for $a=1,2,3$ and $4$ (from bottom to top) for $N=10^4$. Solid lines: simulations. Dashed lines: the leading-order predictions~\eqref{threeeq}. Dash-dotted lines for $a=2,3$ and $4$: more accurate predictions from Eqs.~\eqref{rxt}  and~\eqref{Ma:def} with numerically found $\mathcal{R}(\xi)$.}
\label{moments}
\end{figure}

Figure \ref{moments} shows simulation results for the moments $M_a$. As one can see, $M_1$ approaches the constant value, predicted by Eq.~\eqref{Mastatic}. The growth of $M_2$ with time agrees very well with the prediction from the hydrodynamic theory, and also agrees with the leading-order asymptotic \eqref{logscaling} up to a constant shift $O(1)$ which is beyond the logarithmic accuracy of Eq.~\eqref{logscaling}. $M_3(t)$ and $M_4(t)$ follow the hydrodynamic theory until $t \simeq 500$, where their growth starts to saturate because of the finite $N$.

\section{Finite-$N$ effects}
\label{finiteN}

At large but finite $N$, two types of deviations from the predictions of hydrodynamic theory appear. First, the hydrodynamic predictions break down in the region where there are few or no particles. Second, the swarm fluctuates.  The former effect defines a finite swarm radius, $\ell$, formally defined as the maximum distance of $N-1$ particles (the leader excluded) from the origin. The average value of $\ell$ can be estimated from the simple condition $\int_{\bar{\ell}(t)}^\infty dx\, \rho(x,t) \sim 1/N$,
or equivalently
\begin{equation}
\label{crit}
r[\bar{\ell}(t),t]\sim 1/N\,,
\end{equation}
Of most interest is a late-time regime, $t\gg 1$, which includes two different asymptotic sub-regimes, determined by the interplay between two large parameters, $t\gg 1$ and $N\gg 1$.  For very late times (the condition will be presented shortly) the halo no longer exists because of the particle depletion, and the whole swarm  reaches the steady state described by Eq.~\eqref{rho-2:sol}. In this regime $\bar{\ell}$ is already independent of time, and Eqs.~(\ref{rho-2:sol}) and (\ref{crit}) yield
\begin{equation}
\label{LN:2}
\bar{\ell}\simeq C_2 \sqrt{N}\,.
\end{equation}
where $C_2 \simeq 2.3$ as we obtained in Monte-Carlo simulations.

At earlier times (but still $t \gg 1$) the halo of the swarm is still at work.  Here we can substitute Eq.~(\ref{rxt}) in Eq.~\eqref{crit} and use the large-$x$ asymptotic $r_0(x) \sim x^{-2}$ of Eq.~(\ref{r2:sol}). If the resulting $\bar{\ell}(t)$ is much larger than $\sqrt{4t}$, we can also use  the large-$\xi$ asymptotic (\ref{R:largesol}) of $\mathcal{R}(\xi)$. This calculation yields an algebraic equation for $\bar{\ell}(t)$,
\begin{equation}\label{condell}
\bar{\ell}^3 \exp\left(\frac{\bar{\ell}^2}{4t}\right) = O(N \sqrt{t})\,,
\end{equation}
Its solution can be written as
\begin{equation}\label{prodlog}
\bar{\ell}(t) \simeq \left[6t\,W\left(C_1 N^{2/3} t^{1/3}\right)\right]^{1/2}\,,
\end{equation}
where $W(z)$ is the product log (or Lambert $W$) function
\cite{Lambert}, and $C_1=O(1)$ is an unknown numerical factor. The leading-order asymptotic of the Lambert function $W(z)$ at $z\to \infty$ is $\ln z$, so the Lambert function describes logarithmic dependence of $\bar{\ell}(t)$ on $N$ and determines logarithmic corrections to the simple diffusive scaling $t^{1/2}$.    The large logarithmic factors justify \textit{a posteriori} our assumption that $\bar{\ell}/\sqrt{4t} \gg 1$.

\begin{figure}[ht]
\includegraphics[width=0.46\textwidth,clip=]{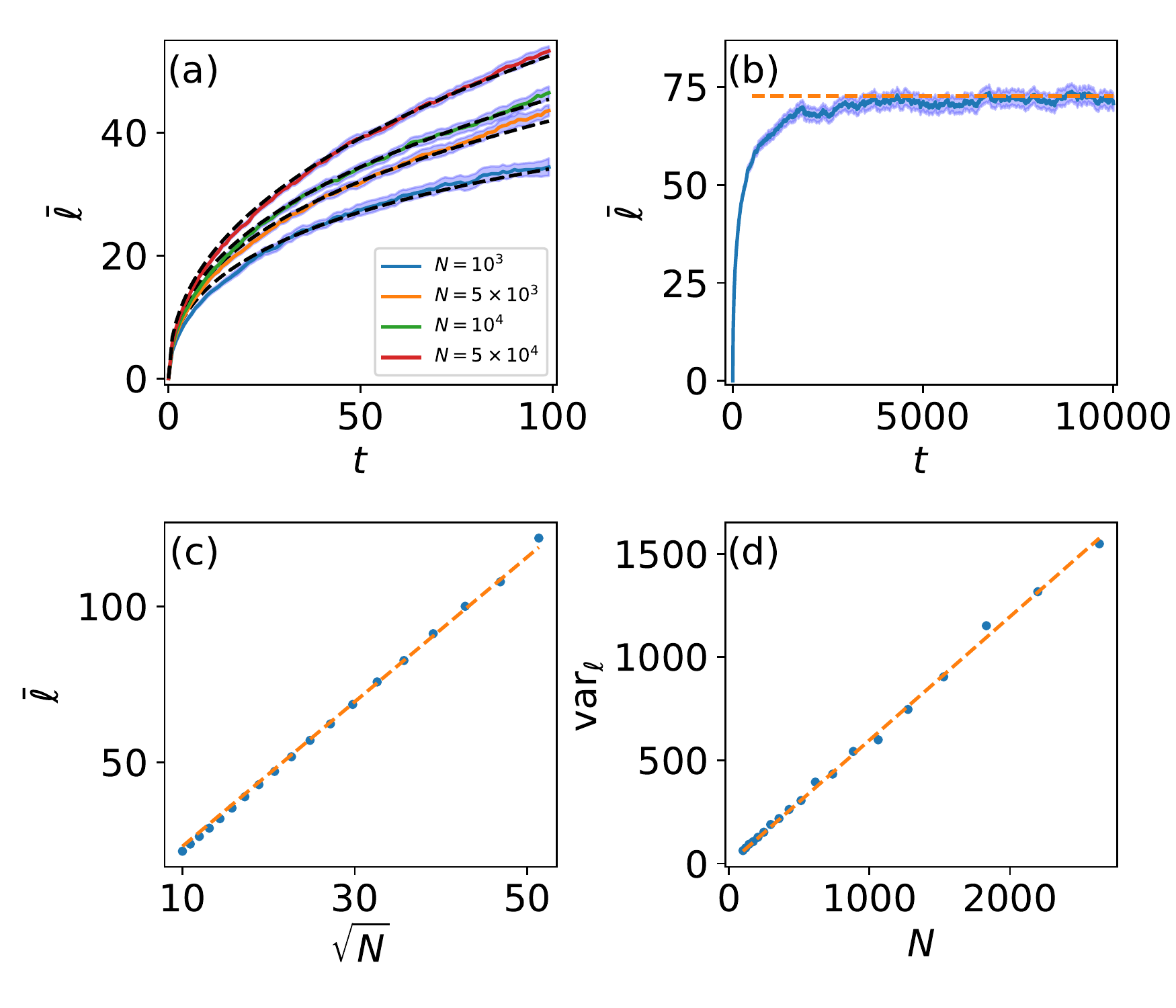}
\caption{(a) The average swarm radius $\bar{\ell}(t)$ versus time for different $N$ (see legend). Solid lines: simulations. Dashed lines: predictions from numerically solving $r[\bar{\ell}(t),t]=1/N$ [Eq.~\eqref{crit}].  (b) Saturation of the growth of $\bar{\ell}$ with time for $N=10^3$. Shaded areas in panels (a) and (b) represent a 95\% confidence interval around the mean. (c) The steady-state value of $\bar{\ell}$ versus $N$. Symbols: simulations.
Dashed line: Eq.~\eqref{LN:2} with $C_2=2.3$. (d) The variance of the swarm radius $\text{var}_{\ell}$ versus $N$ in the steady state.  Symbols: simulations. Dashed line: $\text{var}_{\ell} = B N$, where $B\simeq 0.53$.}
\label{ellMC}
\end{figure}

Equation~(\ref{prodlog}) breaks down at later times,  when $\bar{\ell}(t)$ becomes comparable with the asymptotic time-independent result (\ref{LN:2}). This happens at $t\sim N/\ln N$.   Altogether, our predictions for $\bar{\ell}(t)$ are the following:
\begin{equation}
\label{tworegimes}
\bar{\ell}(t) \simeq
\begin{cases}
 \text{Eq.} \,(\ref{prodlog})\,,& 1\ll t\ll \frac{N}{\ln N}\,,\\
\text{Eq.} \,(\ref{LN:2})\,, & t\gg \frac{N}{\ln N}\,,
\end{cases}
\end{equation}
and it is also assumed that both $\ln t$ and $\ln N$ are very large. It is impractical, however, to meet the latter conditions in Monte-Carlo simulations. Therefore, for moderately large $t$ and $N$, we estimated $\bar{\ell}(t)$ using Eq.~\eqref{crit} with the numerically found $\mathcal{R}(\xi)$ and solved numerically the resulting algebraic equation. The resulting  $\bar{\ell}(t)$ for different $N$ is shown in Fig. \ref{ellMC}(a), and a good agreement between the theory and simulations is observed. Figure \ref{ellMC}(b) shows the growth of $\bar{\ell}$ with time, followed by a saturation predicted by Eq.~\eqref{LN:2}. Figure \ref{ellMC}(c) verifies the predicted $\sqrt{N}$ dependence of $\bar{\ell}$ in the steady state, see Eq.~\eqref{LN:2}.

Another important type of finite-$N$ effects is fluctuations around the hydrodynamic steady state, caused by the discreteness of particles and by the random character of the elemental processes of Brownian motion and competition. One interesting question here concerns fluctuations of the swarm radius $\ell$.  At very long times, not only the average radius $\bar{\ell}$, but the whole $\ell$-distribution approaches a steady state.
Our simulations show [see Fig.~\ref{ellMC}(d)], that  the variance $\text{var}_{\ell}$ of the swarm radius in the ultimate steady state scales linearly with $N$ at large $N$: $\text{var}_{\ell} =B N$, where $B \simeq 0.53$. As a result, the relative magnitude of the fluctuations, $\bar{\ell}/\sqrt{\text{var}_{\ell}}$, is independent of $N$, so the swarm's radius is not a self-averaging quantity.

\section{Dynamics of the leader}
\label{leader}

\begin{figure}[t]
\includegraphics[width=0.5\textwidth,clip=]{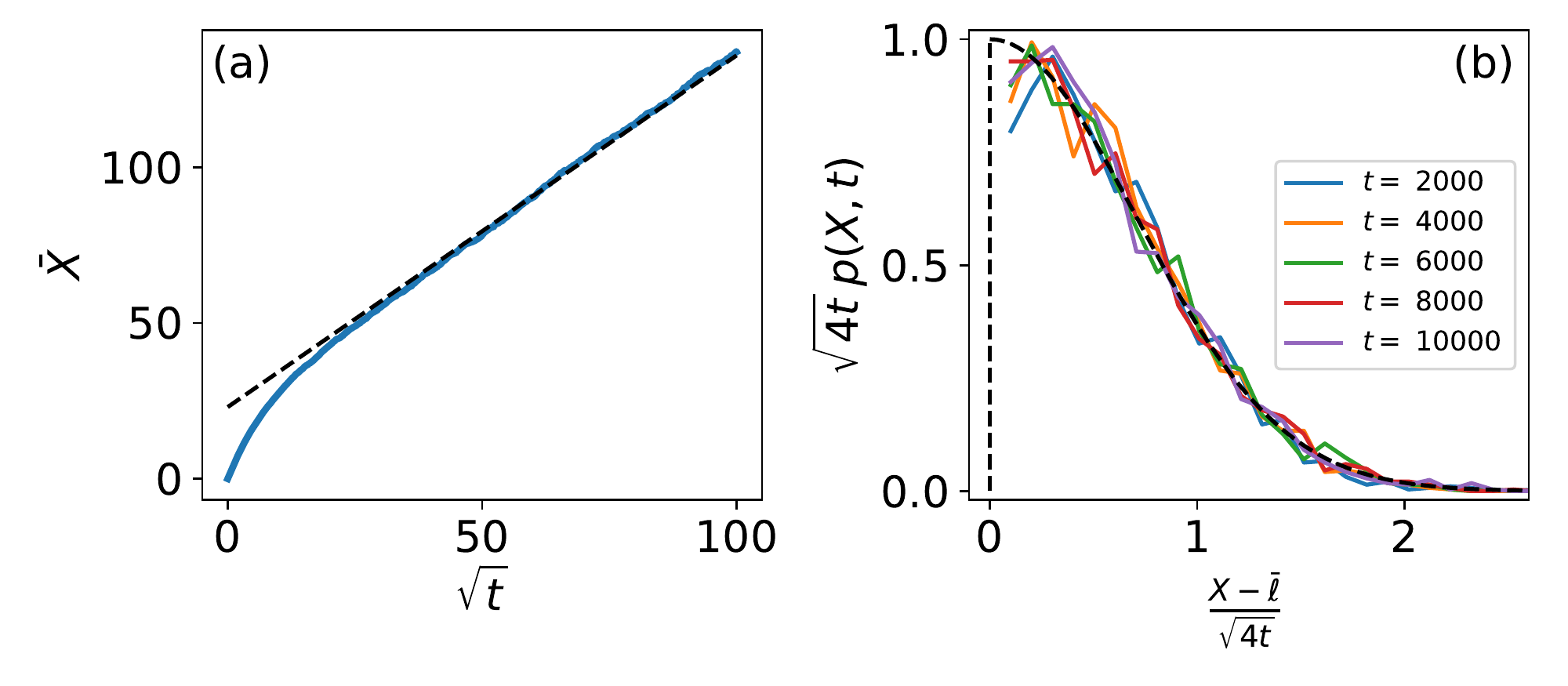}
\caption{(a) The average distance of the leader from the origin vs. time. Dashed line: Eq.~\eqref{avleader}, solid line: simulations. (b) The rescaled distribution of the distance of the leader from the origin at different times. Solid lines: simulations at different times, dashed line: Eq.~\eqref{distleader}. In both panels $N = 10^2$.}
\label{leaderstat}
\end{figure}

By definition, a particle that is currently farthest from the origin  will always win
when competing with any other particle. As a result, the dynamics of the current leader (whose identity can change during the process) is very different from that of the rest of particles: the leader  continues its exploration indefinitely. We argue that the distribution $p(X,t)$ of
the typical fluctuations of the distance $X$ of the leader from the origin follows a simple scenario, where (i) the leader performs  a pure Brownian motion on the ray $\ell(t)<X<\infty$, and (ii) there is a reflecting wall at $X=\bar{\ell}(t)$, where the leader can change its identity.   At very long times $\bar{\ell} =O(N)$ ceases to depend on time, see Eq.~\eqref{LN:2}. As a result, at these long times $p(X,t)$ can be described by the long-time asymptotic solution of the diffusion equation
\begin{equation}\label{diffusion}
p_t = p_{XX}\,,\quad X>\bar{\ell}\,,
\end{equation}
subject to the reflecting boundary condition $p_X(X=\ell,t)=0$. This asymptotic is elementary:
\begin{equation}\label{distleader}
p(X,t) = \frac{e^{-\frac{\left(X-\bar{\ell}\right)^2
   }{4 t}}}{\sqrt{\pi t }}\, \,,\quad X>\bar{\ell}.
\end{equation}
In particular, the average value of $X$ grows as
\begin{equation}\label{avleader}
\bar{X}(t) \simeq\frac{2 \sqrt{t}}{\sqrt{\pi }}+\bar{\ell}\,.
\end{equation}
Equation \eqref{avleader} at long times is compared with simulation results in Fig.~\ref{leaderstat}(a), and a very good agreement is observed.  Figure \ref{leaderstat}(b) compares, at different times, the rescaled distribution $\sqrt{\pi t}\, p(X,t)$ versus $(X-\bar{\ell})/\sqrt{t}$, measured in the simulations,  with theoretical prediction~\eqref{distleader}. Some disagreement at small distances is to be expected as we assumed that the reflecting wall is fixed at $X=\bar{\ell}$, whereas the actual swarm radius exhibits relatively large fluctuations, see Sec. \ref{finiteN}.

\section{Many-particle competition}
\label{sec:n}

\begin{figure}[t]
\includegraphics[width=0.35\textwidth,clip=]{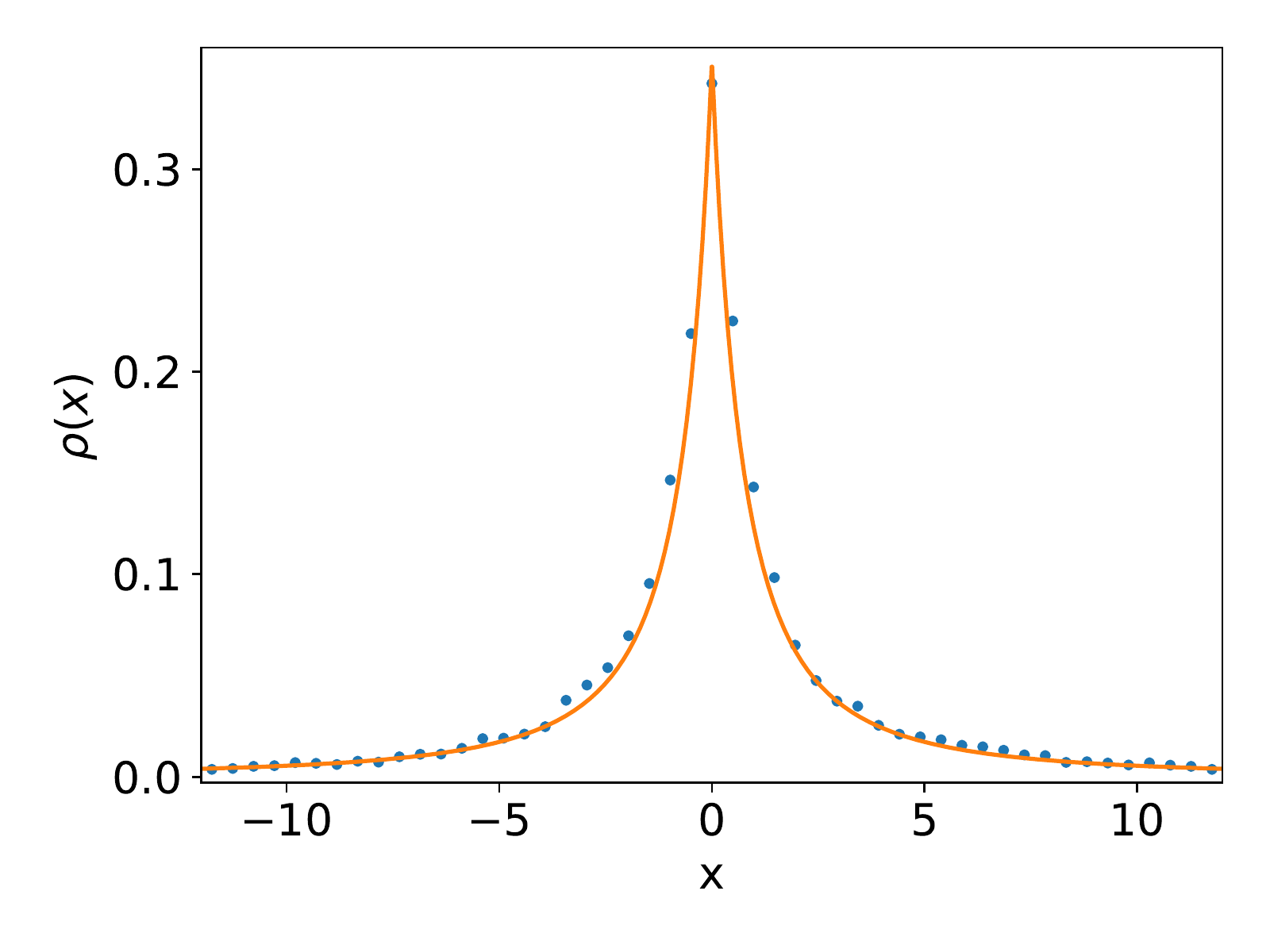}
\caption{The steady state density profile for $n=3$ and $N=10^2$, as obtained in Monte-Carlo simulations  (points) and predicted by Eq.~\eqref{rho-n:sol} (dashed line).}
\label{fig:rho3}
\end{figure}

The inter-particle competition can be readily extended to an arbitrary number of Brownian particles $2\leq n\ll N$.
In the generalized model, in each resetting event, $n$ particles are selected randomly, and the particle closest  to the origin is reset to the origin. One immediate consequence of this rule is the presence of $n-1$ leaders which never lose in the competition and should therefore  be described separately. We will deal with these particles below. The rest of particles can be described by the hydrodynamic theory which generalizes the theory presented in Secs. \ref{sec:2} and \ref{relaxation}. In particular, Eq.~\eqref{eqrho} for the coarse-grained particle density gives way to the following equation:
\begin{equation}
\label{rho-n}
\rho_t = \rho_{xx}-2^{n-1} n \rho r^{n-1}\,,\quad x>0.
\end{equation}
As before, $r=r(x,t)$ is defined by Eq.~\eqref{r:def}, and we have assumed a symmetric initial density profile, $\rho(x,t=0) = \rho (-x,t=0)$. Using Eq.~\eqref{rho-n}, Eq.~\eqref{r-2} is now replaced by the equation
\begin{equation}
\label{r-n}
r_t = r_{xx}-2^{n-1} r^n\,,
\end{equation}
which can be interpreted as a reaction-diffusion equation with a loss reaction of order $n$.
The boundary conditions~\eqref{r:BC} continue to hold.
The hydrodynamic steady state obeys the equation
\begin{equation}
\label{r-n-SS}
r'' =  2^{n-1} r^n\,,
\end{equation}
which can be integrated once to give
\begin{equation}
\label{rn:eq}
r' = -\sqrt{\frac{2^n}{n+1}}\,r^\frac{n+1}{2}\,,\quad x>0\,.
\end{equation}
Integrating Eq.~\eqref{rn:eq} subject to $r(0)=1/2$ [see Eq.~\eqref{r:BC}], we obtain the steady-state profile of $r(x)$:
\begin{equation}
\label{rn:sol}
r_0(x) = \tfrac{1}{2}[1+(n-1)C_n |x|]^{-\frac{2}{n-1}}\,,
\end{equation}
where $C_n=[2(n+1)]^{-1/2}$.
The steady-state density,
\begin{equation}
\label{rho-n:sol}
\rho_0(x) = C_n \left(1+(n-1)C_n |x|\right)^{-\frac{n+1}{n-1}}\,,
\end{equation}
agrees well with our simulations for $n=3$, see Fig. \ref{fig:rho3}. Notice that the as $n$ increases, the tails of the steady-state density become fatter and fatter. The zeroth moment of the density, $M_0$, is convergent for any $n$, but the first moment $M_1$ diverges already for $n=3$. In general, the moment $M_a$ of the steady-state density profile diverges for $a\geq 2/(n-1)$.

When $N$ is finite, the long-time behavior of the density moments $M_a$ for $n>2$ is once more determined by the competition of two large parameters: $t$ and $N$. As in the case of $n=2$, for sufficiently large $N$ there is a dynamical stage where the scaling ansatz~\eqref{rxt} holds, except that the stationary factor $r(x)$ is now given by \eqref{rn:sol}. Inserting the ansatz  \eqref{rxt} into \eqref{r-n}, we obtain an ordinary differential equation for the scaling function:
\begin{equation*}
\label{Rn:eq}
\mathcal{R}''(\xi)\!+\!\left(2\xi\!-\!\frac{4}{(n-1)\xi}\right)\mathcal{R}'(\xi)\!+\!\frac{2(n+1)}{(n-1)^2}\,
\frac{\mathcal{R}\!-\!\mathcal{R}^n}{\xi^2}=0.
\end{equation*}
A numerical solution of this equation with the boundary conditions \eqref{R:BC} enables one
to determine, for $a>2/(n-1)$, the amplitudes of the time-dependent density moments $M_a(t)$ which exhibit  power-law scaling with time.  Notice that for $n=3$ it is the first moment $M_1(t)$ which grows with time logarithmically.

The ultimate steady-state at $t\to \infty$ is still determined by
finite-$N$ effects. In particular, using the condition $r(\bar{\ell})\sim N^{-1}$, we obtain an estimate for the average swarm radius (the maximum distance from the origin of the $N-n+1$ particles who make the swarm) at $t \to \infty$:
\begin{equation}
\label{LN:n}
\bar{\ell} \sim N^\frac{n-1}{2}\,.
\end{equation}
For $n=3$ this gives $\bar{\ell} \sim N$.

\begin{figure}[t]
\includegraphics[width=0.35\textwidth,clip=]{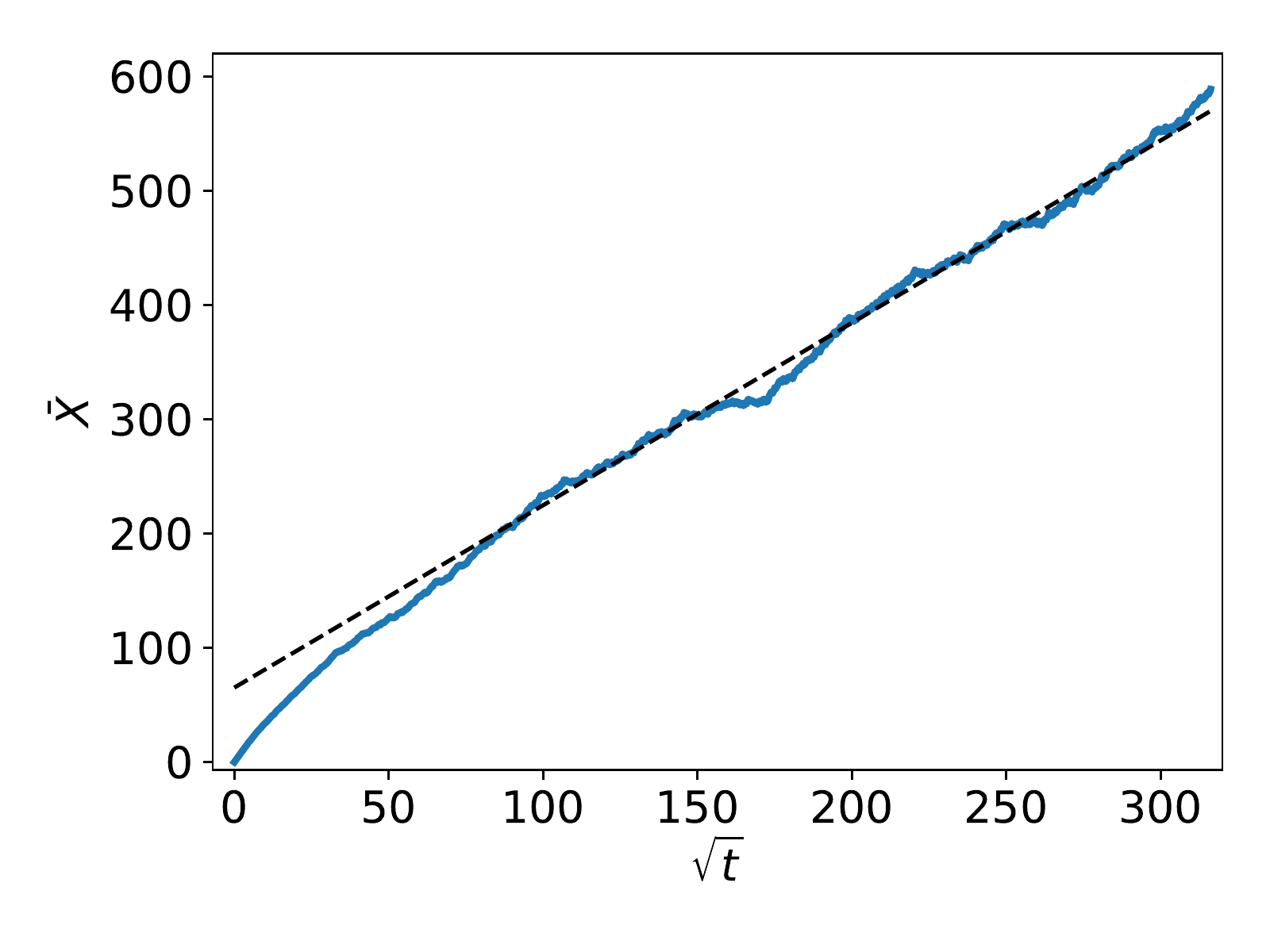}
\caption{The average distance of the leader from the origin vs. time for $n=3$ and $N = 10^2$. Solid lines: simulations, dashed line: $\bar{X}=\mu_3 \sqrt{t}+\bar{\ell}$.}
\label{leaderaverage3}
\end{figure}

Now let us focus on the $n-1$ leaders who ``play a different game" by continuing their eternal Brownian exploration. As they are independent of each other, the joint probability distribution $P(X_1, X_2,\dots X_{n-1},t)$ of their distances from the origin is equal to the product of the single-particle distributions. According to our simple scenario, at long time, each of these single-particle distributions is described by Eq.~\eqref{distleader}. The joint distribution is therefore
\begin{equation*}
\label{joint}
P(X_1, \dots X_{n-1},t) = (\pi t)^{-\frac{n-1}{2}}\,\exp \left[-\sum_{i=1}^{n-1} \frac{(X_i-\bar{\ell})^2}{4t}\right].
\end{equation*}
The distribution $\mathcal{P}_n(X,t)$ of the distance $X$ from the origin of the absolute leader can be found in a standard way, see \textit{e.g.} Ref. \cite{MajumdarSchehr}:
\begin{equation}
\label{mathcalPgeneral}
\mathcal{P}_n(X,t)=\frac{d}{dX}\left[Q(X,t)\right]^{n-1}\,,
\end{equation}
where $Q(X,t) = \int_{\bar{\ell}}^X dX'\,p(X',t)$.  Using Eq.~\eqref{distleader} and evaluating  $Q(X,t)$, we obtain
\begin{equation}
\label{mathcalP}
\mathcal{P}_n(X,t)=\frac{n-1}{\sqrt{\pi t}}\,\,e^{-\frac{(X-\bar{\ell})^2}{4 t}}
   \left[\text{erf}\left(\frac{X-\bar{\ell}}{
   \sqrt{4t}}\right)\right]^{n-2}\,.
\end{equation}
The average distance $\bar{X}$ grows with time diffusively, namely as $\bar{X} (t) = \mu_n \sqrt{t}+\bar{\ell}$, where
\begin{equation}\label{mun}
\mu_n = \frac{4(n-1)}{\sqrt{\pi}}\,\int_0^{\infty} dz\,z\, e^{-z^2} [\text{erf}(z)]^{n-2}.
\end{equation}
In particular, $\mu_2 = 2/\sqrt{\pi}$ in agreement with Eq.~\eqref{avleader}. Further,
\begin{equation}\label{mu23}
\mu_3=\frac{2\sqrt{2}}{\sqrt{\pi}}\,,\quad \mu_4 = \frac{12 \sqrt{2}}{\pi^{3/2}}\arctan \frac{1}{\sqrt{2}}\,, \quad \dots \,.
\end{equation}
The prediction for $n=3$ is compared with simulations in Fig.~\ref{leaderaverage3}, and a good agreement is observed.

\section{Discussion}
\label{discussion}

We introduced and studied analytically and numerically a simple $N$-particle model which combines Brownian motion with inter-particle competition encouraging achievers.  In the $N\to \infty$ limit, the swarm density follows a nonlocal hydrodynamic theory and ultimately relaxes to a stationary density profile. This profile exhibits a power-law decay at large distances followed by a cutoff at a (fluctuating) distance $O(\sqrt{N})$ for the 2-particle competition, and $O(N^{\frac{n-1}{2}})$ for the $n$-particle competition. At intermediate times, the relaxation process exhibits a
a non-stationary halo in a peripheral region, which expands in a self-similar manner.

We further showed that there are $n-1$ particles (the current leaders) which follow different dynamics: They cannot lose in the competition and continue their Brownian exploration forever. We suggested a simple scenario for typical fluctuations of the leaders, where in particular, the location of the leader grows diffusively with time.

Our results capture some prominent features of the distribution of wealth. Among these are the power law decay of the distribution of incomes \cite{Pareto1897}, and the fact that the exponent of this power law can vary  depending on the number of agents $n$ in each inter-agent competition \cite{billionaires2021}.
Another noteworthy result is the presence of expanding halo, indicating inequality that increases over time \cite{RisingInequality2001}: while the bulk of the distribution (most incomes) is already in a steady state, the distribution tail (extremely large incomes) continues to grow for sufficiently many agents. The presence of individual leaders who ``play a different game" is also a natural consequence of encouraging the achievers. Given these interesting features, our model does seem to capture some basic mechanisms leading to a broad wealth distribution.

On the physics side, future work can deal with extensions of the model to higher dimensions. One can also try to develop a macroscopic fluctuation formalism in the spirit of fluctuating hydrodynamics of Landau and Lifshitz \cite{Spohn}, as it has been recently done for three other $N$-particle models with reset \cite{VAM2022,bees5}. Such a framework should be useful for studying fluctuations of macroscopic quantities: for example, of the center of mass of the swarm. An even more interesting question about the (intrinsically large) fluctuations of the swarm's radius (\textit{cf}. \cite{VAM2022,bees5}) will most likely demand different methods.

\section*{Acknowledgements}
PK is grateful to Satya Majumdar for a useful discussion. BM acknowledges support from the Israel Science Foundation (ISF) through Grant No. 1499/20.

\end{document}